\documentclass[namedreferences]{kluwer}
\usepackage{graphicx}
\def\arcsec{\hbox{$^{\prime\prime}$}} %
\def\deg{$^{\circ}$} %
\begin{document}
\begin{article}
\begin{opening}
\title{ %
The GraF instrument for imaging spectroscopy with the adaptive
optics\footnote{ %
Based on observations collected at European Southern Observatory, Chile, La 
Silla 3.6\,m telescope, ESO period 59, technical time.} %
}
\author{A. \surname{Chalabaev}} %
\author{E. \surname{le Coarer}} %
\author{P. \surname{Rabou}} %
\author{Y. \surname{Magnard}} %
\author{P. \surname{Petmezakis}} %
\institute{Laboratoire d'Astrophysique, Observatoire de Grenoble, UMR 5571,
CNRS and Universit\'{e} Joseph Fourier, B.P. 53X, F-38041 Grenoble, France} %
\author{D. \surname{Le Mignant}} %
\institute{W.\,M.\,Keck Observatory, 65-1120 Mamalahoa Highway, Kamuela HI96743, 
U.S.A. }
\runningtitle{GraF instrument for imaging spectroscopy} \runningauthor{Chalabaev 
et al.}
\begin{ao}
e-mail: Almas.Chalabaev@obs.ujf-grenoble.fr
\end{ao}
\begin{abstract}
The GraF instrument using a Fabry-Perot interferometer cross-dispersed
with a grating was one of the first integral-field and long-slit
spectrographs built for and used with an adaptive optics system. We
describe its concept, design, optimal observational procedures and the
measured performances. The instrument was used in 1997-2001 at the ESO
3.6\,m telescope equipped with ADONIS adaptive optics and SHARPII+
camera. The operating spectral range was 1.2\,-\,2.5 $\mu m$. We used
the spectral resolution from 500 to 10\,000 combined with the angular
resolution of 0.1{\arcsec}\,-\,0.2{\arcsec}. The quality of GraF data
is illustrated by the integral field spectroscopy of the complex
$0.9{\arcsec} \times 0.9{\arcsec}$ central region of $\eta$ Car in the
1.7 $\mu m$ spectral range at the limit of spectral and angular
resolutions.
\end{abstract}
\end{opening}
%
\section{Introduction}\label{intro}
High angular resolution observations at the diffraction limit of the ground 
based large telescopes were pioneered using the speckle analysis of images 
\cite{labeyrie}, however their sensitivity was severely limited by the necessity 
to keep exposures short in order to ``freeze'' the turbulence induced patterns.  
The adaptive optics overcame this obstacle allowing long exposures (COME-ON and 
ADONIS at the ESO 3.6\,m telescope, \opencite{adonis97}, PUEO at the CFHT, 
\opencite{pueo98}, NAOS at the ESO VLT, \opencite{lagrange03}, and others).  The 
limiting magnitude on the telescopes equipped with the adaptive optics 
(hereafter AO) is now defined by the detector and optics efficiency as for any 
observations, while the angular resolution is close, at least in the IR, to the 
telescope diffraction limit, about 0.1{\arcsec} for a 4\,m aperture at $\lambda 
= 2$\,$\mu m$.

The achievements of the AO were firstly exploited for imaging (see reviews by 
\opencite{close00}, \opencite{lai00}, \opencite{menard00}).  The next step was 
to use the adaptive optics for spectroscopy.  Indeed, the combination of high 
angular and high spectral resolutions is a key issue for a number of 
observational programmes such as physics and evolution of multiple stellar 
systems, morphology and dynamics of circumstellar gas and jets in young and 
evolved stars, etc.  

Driven by this ideas, the GraF project was started in 1995 with the aim
to build an imaging spectrograph optimized for the use with the ADONIS
at the ESO 3.6\,m telescope. The instrument is based on the integral
field spectroscopic properties of the Fabry-Perot interferometer
(hereafter FPI) used in cross-dispersion with a grating (le Coarer et
al., 1992, 1993). It was tested at the telescope in 1997 and
successfully used for a number of astronomical programmes in 1998-2001
(for preliminary results see Chalabaev et al., \citeyear{jena99},
\citeyear{graf99}, \citeyear{etacar99}, \opencite{trouboul99}).

In the present article we describe in details the optical concept and
the instrument built to fit the constraints imposed by ADONIS
\cite{adonis97} and the SHARPII+ camera \cite{sharp95}. We give the
account of the observational procedures emerged from our experience as
optimal, present the measured performances, and illustrate them by a
sample of reduced data obtained at the limit of the instrument
possibilities in terms of the angular and spectral resolutions.

The discussion will be restricted to the applications of a high
spectral resolution ($R \simeq 5000$\,-\,$10000$) in a moderate
passband ($\Delta \lambda \leq 50$) combined with a high angular
resolution ($\simeq$ 0.1{\arcsec}\,-\,0.2{\arcsec}) in a small field of
view ($FOV \leq 15{\arcsec}$). The operating spectral range of the
instrument was $1.2\,-\,2.5$ $\mu m$.

We hope that this one of the first experiences of the imaging
spectroscopy with the adaptive optics (see also \opencite{tiger95},
\opencite{lavalley97}) will be useful for designers and users of future
spectro-imaging instruments combining the high angular and high
spectral resolutions.
%
\section{Image restoration aspects}\label{image_restoration}
Let us in what follows to make the emphasis of the discussion on the
scarcely resolved objects, i.e.\ having the spatial\footnote
{We will say indifferently ``spatial'' and ``angular'', the corresponding 
variables on the celestial sphere being related by a simple scaling 
factor.}
spectrum in the Fourier domain comparable in the extent to that of the
AO corrected telescope modulation transfer function (MTF). 

There are two reasons for such a choice. Firstly, as witnessed by the
reviewers (cf.\ \opencite{close00}, \opencite{lai00},
\opencite{menard00}), the AO provides a considerable scientific
contribution mainly in the case where the object under study, having
remained point-like at lesser angular resolutions, becomes scarcely
resolved, revealing new spatial features.

The second reason is dictated simply by the fact that the scarcely
resolved objects are the most difficult to measure, so that the overall
performance of a new instrument at the limit of resolution is best
evaluated on such objects.

We can already note that the importance of the scarcely resolved
objects shapes the specifications on the AO assisted spectrograph.
Indeed, to recover the information up to the highest possible spatial
frequency implies the image restoration, and this aspect has
necessarily to be taken into account in the design of the spectrograph.
\subsection{General}
We shall consider the flux density distribution $S(x, y, \lambda)$
which is non-zero at least at 2 points of the two-dimensional (2D) sky
field \{$x, y$\}. Here, $x$ and $y$ are the spatial coordinates, and
$\lambda$ is the wavelength. As it was already said in
Sect.\,\ref{intro}, the discussion will be restricted to small fields,
$\leq 15{\arcsec}$.

The image restoration problem of imaging spectroscopy in the general
case consists in finding the best estimate ${\hat{S}}(x, y,\lambda)$ of
the object flux distribution density $S(x, y,\lambda)$, satisfying the
tri-dimensional integral equation of convolution:
\begin{equation}  
F(x, y,\lambda) = \int_{}{}\int_{}{}\int_{}{} d\xi d\zeta dw \cdot S(\xi, \zeta, 
w) \cdot {\cal{G}} (\xi - x, \zeta - y, w - \lambda)
\label{eq:inverse3Da}
\end{equation} 
where $F$ is the measured flux density distribution and $\cal{G}$ is the 
instrumental impulse response.

The solution of the integral equation of convolution is known to be
unstable (\opencite{turchin71}; see also \opencite{lucy94review}). The
calibration errors of $\cal{G}$ are strongly amplified in the final
result of image restoration, in particular at high frequencies, so that
the solution $\hat{S}$ has to be regularized, i.e.\ searched in a space
of appropriate smooth functions (see for details \opencite{tikhonov77},
\opencite{titterington85}, \opencite{lucy94strategy}).

The image restoration can be carried out using either one of the many
proposed deconvolution algorithms (e.g.\ \opencite{cornwell92},
\opencite{magain98}, \opencite{lucy03}, and references therein), or by
modeling $\hat{S}$ from \textit{a priori} defined physical
considerations and then searching for the best fit of the convolution
product $\hat{S} \otimes {\cal{G}}$ to $F$ within the physical model.

It is clear that whatever the method adopted for the image restoration,
the accurate calibration of $\cal{G}$ is of high importance. Let us
analyze the sources of the calibrations errors in the case of imaging
spectroscopy in order to get guidelines of the instrumental concept.
\subsection{Calibration errors}
Firstly, we can simplify the Eq.(\ref{eq:inverse3Da}) by noting that
for the considered here angular fields, spectral passbands and
resolutions (see Sect.\,\ref{intro}) the 3D impulse response $\cal{G}$
can be written as the product of the spatial point-spread function
(psf) ${\cal{P}}(x,y |\xi,\zeta; \lambda)$ and the instrumental
spectral profile ${\cal{L}}(x,y; \lambda | w)$ (see e.g.\
\opencite{perina}, \opencite{goodman68}, \opencite{mariotti}). Then we
get:
\begin{equation}  
F(x, y,\lambda) = \int_{}{} dw \cdot {\cal{L}}(x,y,\lambda - w) 
\int_{}{}\int_{}{} d\xi d\zeta \cdot S(\xi, \zeta, w) \cdot {\cal{P}} (\xi - x, 
\zeta - y, \lambda)
\label{eq:inverse3Db}
\end{equation} 

We will further assume that the relevant scientific information can be
extracted without deconvolution on $\lambda$. In other words, we will
limit the discussion to the frequently encountered case where the
instrumental spectral profile ${\cal{L}}$ is much narrower than the
studied spectral lines.

We can then write:
\begin{equation}  
{\cal{L}}(x,y,\lambda - w) \simeq L(x,y,\lambda) \cdot \delta(\lambda-w)
\label{eq:mdelta}
\end{equation} 
where $L(x,y,\lambda)$ is the spectral transmission, a simple 
multiplicative factor, and $\delta(\lambda-w)$ is the Dirac impulse 
function. 

The convolution concerns now only the spatial dimensions, so that the
Eq.(\ref{eq:inverse3Db}) is further simplified as follows:
\begin{equation}  
F(x, y,\lambda) = L(x,y,\lambda) \int_{}{}\int_{}{} d\xi d\zeta \cdot S(\xi, 
\zeta, \lambda) \cdot {\cal{P}} (\xi - x, \zeta - y, \lambda)
\label{eq:inverse2D}
\end{equation} 
The important consequence is that in the absence of the deconvolution
on $\lambda$ the contribution of the calibration errors on ${\cal{L}}$
to the uncertainty of the final estimate $\hat{S}$ is considerably
reduced.

Furthermore, we note that usually the variations of $\cal{L}$ during
observations are due to mechanical flexure at the telescope. They are
slow, with the time scale of tens of minutes, and can be monitored with
a good accuracy. Similarly, the dependence of $\cal{L}$ on $x$ and $y$
is stable and can be calibrated accurately.

In contrast to the relative stability of ${\cal{L}}$, the psf
${\cal{P}}$ undergoes significant temporal variations on the time scale
of minutes or even shorter due to atmospheric turbulence. Although the
AO improves dramatically the situation, the residual wavefront
variations can still be significant. Their amplitude depends on the
operating wavelength, the amplitude and the time scale of the
atmospheric turbulence, the performances of the AO system and the
telescope aperture size. With the ESO 3.6\,m telescope and the ADONIS,
the variations of ${\cal{P}}$ expressed in terms of the Strehl ratio
were found to change from 10 to 35 in the K band for the seeing value
of $\approx 1.5{\arcsec}$ within the time interval of 10\,$s$
(\opencite{adonis_lemignant}). The variability is \textit{a fortiori}
stronger at the shorter wavelengths of J and H bands.

As to the wavelength dependence of ${\cal{P}}$, it is negligible within
the considered here spectral range of a single instrument setting,
$\Delta \lambda/\lambda \leq 50$.

We can conclude that the main source of errors in the final result of
image restoration, the flux distribution estimate $\hat{S}$, comes from
the calibration of the spatial psf ${\cal{P}}$ due to its rapid
variability. The way it is calibrated needs thus special attention. For
instance, the calibration errors can be substantially reduced if
${\cal{P}}$ is measured simultaneously with the measurement of $F$,
e.g.\ on a suitable source in the same field. If ${\cal{P}}$ can be
measured only on a source off the field, it must be done as close in
time as possible. The instrumental concept has to take into account 
these observational aspects.

It also appears important that the two-dimensional $\{x,y\}$-structure
of the field and of the psf $\cal{P}$ is recorded with no scanning%
\footnote{We restrict the term of scanning to the recording data point
after point, or line after line, distinguishing it from the mosaicing,
i.e.\ recording a subregion by a subregion.}
on $\{x,y\}$, in order to keep the error on $\cal{P}$ homogeneous over $x$
and $y$, thus avoiding a random scrambling of spatial features, which
can be unrecoverable.

As to the spectral response $\cal{L}$, the contribution of its
calibration errors in the final uncertainty on $\hat{S}$ is
considerably lesser than that of $\cal{P}$. Indeed, the instrumental
spectral profile $\cal{L}$ is relatively stable and can be calibrated
much more accurately than ${\cal{P}}$. Furthermore, if no deconvolution
is justified on $\lambda$, i.e. if the spectral resolutions high
enough, then the error on $\cal{L}$ propagates into the error on
$\hat{S}$ without amplification.
\subsection{Implications for the AO assisted spectroscopy}
\label{implications}
The above given analysis provides the guidelines of the instrumental
concept for the AO assisted imaging spectroscopy which can be briefly
summarized as follows: (i) the calibration of the psf $\cal{P}$ has to
be simultaneous, or as close in time as possible, to the measurement of
the flux distribution $F$; (ii) the scanning over the spatial $x$- and
$y$-axes must be avoided; (iii) if scanning is unavoidable due to the 
volume of data to be recorded, the instrument concept allowing 
$\lambda$-scanning is preferable.

Obviously, the ideal spectro-imaging instrument would record the entire
cube of data $F(x,y,\lambda)$ in one single exposure with no scanning,
satisfying the criteria of both the accurate image restoration and the
saving the telescope time.

The work is in progress on the ``3D~detectors'' able to record both
the position ${x, y}$ and the energy $\lambda$ of the photons,
\textit{cf.}\ the superconducting tunnel junctions (\opencite{stj94},
\opencite{stj00}) or the dye-doped polymers (\opencite{shbd95}). 
However, their sensitivity is still less convincing than that of the
modern 2D-detectors.

The best available today solutions are offered by the optical set-ups
known as the integral field spectrographs (IFS, \opencite{courtes82},
\opencite{tiger95}, \opencite{mpi3d94}, \opencite{vimos98},
\opencite{spiffi00}). They use the 2D detectors to record
$F(x,y,\lambda)$ with no scanning, provided the detector size is large
enough to record the 3D data cube at once.
%
\section{Particular case of a ``linear image''}
\label{linear}
%
Let us also consider the simplest case of the imaging spectroscopy when
the object flux density $S(x, y, \lambda)$ is non-zero along a straight
line, so that $S$ is a function of only two variables, $S(x, \lambda)$.

In this ``linear'' case, frequently encountered in the astrophysical practice 
(e.g.\ binary stellar systems), the common grating spectrographs recording 
$F(x,\lambda)$ at a single setting offer a suitable solution, insuring the 
calibration of psf ${\cal{P}}$ homogeneous over the studied field \{$x$\}.  
Furthermore, in the case of a circumstellar nebular object, when the extended 
feature is a gas emitting only in spectral lines, the emission in the continuum 
corresponds to the point-like star and provides the calibration of 
${\cal{P}}(x)$ simultaneous to the measurement of $F$.
%
\section{The minimum 3D data cube volume}
\label{data-volume}
%
At high angular and high spectral resolutions, the volume of data to
be recorded in one observation of imaging spectroscopy can be large. 
Let us to estimate what is its minimum under typical astrophysical
specifications.

Empirically, the angular field of $\simeq 3{\arcsec} \times 3{\arcsec}$ would be 
adequate to study most types of objects of interest of the stellar physics.  The 
pixel size of 0.05{\arcsec} is fixed by the angular resolution, which is 
$\approx 0.1{\arcsec}$ at a 4\,m telescope at $\lambda=2\,\mu$m.  Thus, the 
record of the spatial flux distribution at a given wavelength $F_{\lambda}(x, 
y)$ consists of $60\times 60 = 3600$ data values.  Along the $\lambda$-axis, the 
minimum of about 100 points is suitable in order to record the profile of a 
spectral line and the adjacent continuum.  Thus, the entire cube data volume is 
at least $60\times 60 \times 100 \simeq 600^{2}$ values.

This largely exceeds the $256^{2}$ size of the detector available with
ADONIS/SHARPII+. In our case the IFS approach is clearly unaffordable,
the scanning is imposed. Then, in agreement with the conclusions of
error analysis in imaging spectroscopy given in
Sect.\,\ref{implications}, we adopted the concept of the
$\lambda$-scanning imaging spectrograph described below.
%
\section {The GraF concept}
%
\subsection{Optical scheme.  Data cube structure}
\begin{figure} 
	 \centering
	{\includegraphics[width=9.5cm]{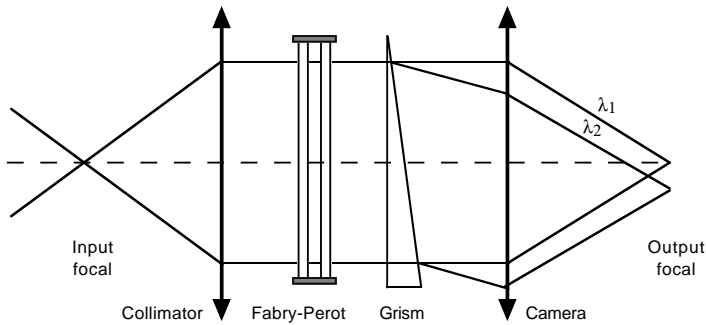}} %
	\caption{The optical concept of the the imaging spectrograph using a 
	Fabry-Perot interferometer in cross-dispersion with a grating, replaced here 
	by a grism for the convenience of the drawing.  }
	\label{fig:graf_concept}
\end{figure}

The concept uses a Fabry-Perot interferometer in cross-dispersion with a grating 
(hence the acronym we gave for this concept, GraF = \textbf{Gra}(\textit{ting}) 
and \textbf{F}(\textit{abry-Perot}).  The set-up (see 
Fig.\,\ref{fig:graf_concept}) was first described by \inlinecite{fabry}, and 
used for spectroscopy by \inlinecite{chabal} and \inlinecite{kulagin80}.  The 
IFS property of the set-up was noticed by \inlinecite{elc-phd} and demonstrated 
by le Coarer et al.\ (\citeyear{morgane92}, \citeyear{morgane93}).  
\inlinecite{baldry} went further, describing a tunable echelle imager, where a 
FPI is cross-dispersed with a grism and with an echelle grating.

In a single frame, GraF records the quasi-monochromatic images of the field 
$F_{\lambda}(x,y)$ corresponding to several values of $\lambda$, which is the 
distinctive property of an IFS instrument.  The spectrum is sampled by a comb of 
the FPI transmission peaks (interference orders) separated by the interorder 
wavelength spacing $\Delta \lambda_{f}$ (Sect.\,\ref{theory}).  The whole set of 
$\lambda$ values is recovered by scanning.
%

\begin{figure} 
	 \centering
	{\includegraphics[width=11.5cm]{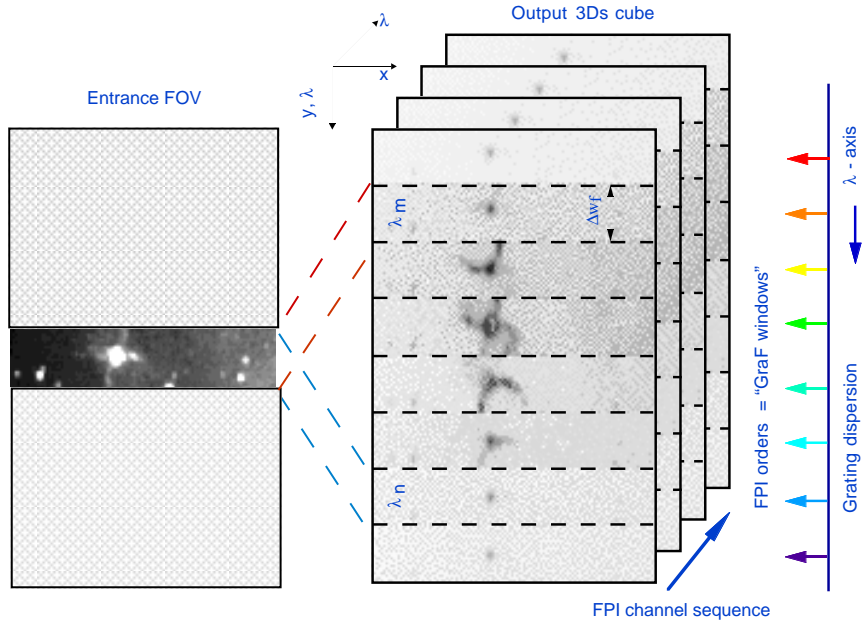}} %
	\caption{The structure of the GraF IFS cube.  \textit{Left}: The ``Southern 
	Crab'' planetary nebula as it appears in the white light (simulated).  The 
	field of view is limited by a rectangular entrance aperture.  
	\textit{Right}: The images of the nebula in the light of a atomic spectral 
	line (simulated image) as they appear in the focal plane of the GraF 
	instrument.  Each frame consists of several monochromatic ``windows'' 
	corresponding to the FPI orders.  Due to the differential motion of the 
	nebulae, the aspect of the nebula is changing from one ``window'' to 
	another.}
	\label{fig:crabe}
\end{figure}

The structure of the spectro-imaging data cube is illustrated in 
Fig.\,\ref{fig:crabe}.  The FPI acts as a ``multi-passband'' filter.  The light 
of different FPI orders is sorted by the grating according their wavelength.  
The resulting series of quasi-monochromatic images of the sky field is formed in 
the output focal plane.  Note that the width\footnote{%
As in the slit spectroscopy, the width and the height are respectively
the directions along and across the grating dispersion. At the same
time, these directions correspond to the spatial axes which will be
denoted hereafter as respectively $x$- and $y$-axes.%
} of the entrance field must be 
limited by a focal aperture to avoid superposition of order images.

Anticipating a detailed discussion (Sect.\,\ref{theory}), let us give the 
figures of a typical cube volume recorded with the actually built instrument.  
At each step of scanning, the detector frame of $256^{2}$ pixels records 8 
narrow-band images corresponding to the dispersed sequence of the FPI orders as 
selected by the grating angle value.  Each FPI ``order image'' covers the same 
1.5{\arcsec}$\times$12.4{\arcsec} region of the entrance sky field sampled with 
the pixels of 0.05{\arcsec}, which makes $\simeq 30 \times 250 = 7440$ spatial 
pixels.  

This is close to the estimated minimum required by the stellar observations, 
although, the width of the FOV $\simeq$ 1.5{\arcsec} is a factor of 2 less than 
the desirable $\simeq 3{\arcsec}$ value.  For the observations of the elongated 
objects, this shortage is partially compensated by the considerable height of 
the FOV of $\simeq 12{\arcsec}$, which can be suitably aligned.

The spectral band covered by the detector is about 40 nm at $\lambda = 2.2$ 
$\mu m $, corresponding to 8 FPI orders.  It is scanned in 48 FPI channel 
frames\footnote
{The spectral step between FPI samples, $h$, is smaller than $FWHM$/2 often 
quoted as the ``Nyquist'' sampling.  We will comment on this issue in Sect.\ 
\ref{FPI_formulae}.}, although in special cases one can limit the scan to a 
narrower range of interest.
The number of spectral samples in a full FPI scan is $48 \times 8 = 384$.  The 
spectral passband of an image is $\delta \lambda \simeq 0.3$\,nm at 
$\lambda$=2\,$\mu m $, so that the corresponding spectral resolving power is $R = 
\lambda/\delta \lambda \simeq 7000$.  The total volume of the cube is $48\times 
256^{2} = 3.15 \cdot 10^{6}$ pixels, or $\simeq 1770^{2}$ pixels.
%
\subsection{Advantages and limitations of the concept}

The GraF $\lambda$-scanning IFS appears as a suitable solution when
scanning is imposed by the modest detector size. It allows a
simultaneous record of several monochromatic images of a reasonably
large field thus keeping possible an accurate image restoration.

Scanning is done by the comb of spaced FPI transmission peaks rather
than by a contiguous set of the wavelength values like in a grating
based instrument. The peaks spacing makes more certain to have in each
frame at least one ``order'' image placed at the wavelengths of the
continuum emission, thus providing a reference signal for photometric
monitoring and, in the nebular cases, the calibration of the
${\cal{P}}$ simultaneous to the measurement of $F$.

Another convenient point of the concept is the simplicity of transforming a 
grating spectrograph into a GraF instrument by adding solely a FPI and an 
adjustable slit (see also comments by \opencite{baldry}).  \textit{Vice versa}, 
the instrument is easily switched to a grating spectrograph for observations of 
``linear'' objects.

However, the GraF concept is intrinsically scanning, so that if
scanning is unnecessary, the GraF is slower than other mentioned above
IFS concepts.

Further, the width of the FOV has an upper limit. Expressed in the
elements of the angular resolution $\delta \phi$, the FOV width cannot
exceed the FPI finesse value $\cal F$ (see Sect.\,
\ref{graf_formulae}). For the maximum $\cal F$ of $\simeq 40$ still
suitable for the high-throughput imaging (\opencite{bland95}), the
maximum FOV width of a GraF is $\simeq 4{\arcsec}$ for $\delta \phi
\simeq 0.1{\arcsec}$.
%
\section{Formulae for GraF optical parameters}\label{theory}
\subsection{FPI formulae}\label{FPI_formulae}
Let us remind the basic terms describing the FPI properties.  The FPI 
spectral transmission is a comb of peaks, called ``orders'', occurring at 
$\lambda_{m}$ defined by the condition of the interference:
\begin{equation} 
m \lambda_{m} = 2 n e \cos i
\label{eq:basicFP}
\end{equation} 
where the integer $m$ is the order of the interference, $n$ the 
refractive index, $e$ the gap between the FPI plates, and $i$ the angle 
of incidence.

The value of $\lambda_{m}$ varies over the FOV according to the value
of $i$.  However, as it can be estimated from Eq.\,\ref{eq:basicFP},
this variation is negligible: ${\delta \lambda}/{\lambda} \simeq
5\cdot 10^{-9}$ for the considered fields of view $\leq$~15{\arcsec}. 
In what follows, it will be assumed that $i = 90$\deg, so that the gap
$e$ defines completely the set of $\lambda_{m}$.

The distance between two neighbor orders is called the interorder spectral 
spacing.  Expressed in the wavelength units, $\Delta \lambda_{f}$, it can 
be written from Eq.\,\ref{eq:basicFP} as follows:
\begin{equation} 
\Delta \lambda_{f} = \frac{\lambda_{m} \lambda_{m+1}}{2 n e} \simeq 
\frac{\lambda^{2}}{2 n e}
\label{eq:interorder}
\end{equation} 

It is often more convenient to use the interorder spacing expressed in the 
wavenumber units $\Delta \sigma_{f}$, which has no spectral dependence:
\begin{equation} 
\Delta \sigma_{f} = \frac{1}{2 n e}
\end{equation} 

The element of spectral resolution $\delta \lambda$ will be defined as
the $FWHM$ of the instrumental profile $\cal{L}$.  It has the
advantage to be immediately measurable (see however
\opencite{jones_bland95} for different criteria of resolution).

The effective finesse $\cal{F}$ is defined as the ratio of the interorder 
spectral spacing to the element of spectral resolution:
\begin{equation} 
{\cal F} = \frac {\Delta \lambda_{f} }{\delta \lambda}
\label{eq:finesse} 
\end{equation}
The instrumental profile $\cal{L}$ is the Airy function:
\begin{equation} 
{\cal{L}} = \frac{1}{1+{\cal{F}}sin^{2}(\pi \sigma/\Delta \sigma_{f}}
\end{equation} 
For high values of $\cal{F}$, it can be approximated by the Lorentzian:
\begin{equation} 
{\cal{L}} = \frac{1}{1+(2 \sigma/\delta \sigma)^{2}}
\end{equation} 

Using Eqs.(\ref{eq:interorder}) and (\ref{eq:finesse}), the spectral resolving 
power $R = \lambda/\delta \lambda$ can be formally written as follows:
\begin{equation} 
R = \frac{2 n e \cal F}{\lambda}
\end{equation} 

Given that the finesse $\cal F$ approximately corresponds to the number of the 
effective reflections on the FPI plates, the maximum optical path difference 
between the interfering wavefronts $OPD_{max}$ can be written as follows:
\begin{equation} 
OPD_{max} \approx 2 n e {\cal{F}}
\label{eq:dmax}
\end{equation} 

The sampling step we used for the FPI scanning is $h = \delta \lambda
/2.7$.  This appears as ``oversampling'' as compared to the
``canonic'' value of $h_{s} \simeq 2$.

There are two ways to justify the used value of $h$. The first one is
to argue that the Airy profile of $\cal{L}$ has a considerably sharper
core than the more familiar Gaussian or $sinc^{2}$ profiles of grating
spectrographs, so that the rate based on the convention $\delta \lambda
= FWHM$ would certainly undersample the core profile.

More strictly speaking, let us recall that the step $h$ allowing to
fully interpolate the function with a finite Fourier spectrum must be
$h \leq h_{s} = 1/(2\cdot f_{cut})$ (Shannon theorem), where $f_{cut}$
is the cut-off frequency of the Fourier spectrum. It happens that for a
Gaussian profile, $\delta \lambda = FWHM \simeq 1/f_{cut}$, which
conveniently translates into ``the rule'' $h_{s} = \frac{1}{2} \delta
\lambda$. The trouble with FPI is that the Fourier transform of
$\cal{L}$ is never zero\footnote%
{\textit{cf}.\ the Fourier transform of a Lorentzian is $exp(-\delta
\sigma \pi f)$. }
and has no $f_{cut}$. Thus, formally FPI escapes the Shannon theorem.

In practice, the effective number of reflexions on the FPI plates,
hence the effective finesse $\cal{F}$ and the effective $OPD_{max}$, is
limited by the noise, so that $f_{cut}$ can be defined on this basis.
For the same instrument, it will vary depending on the S/N ratio. The
uncertainty on $f_{cut}$, and hence on $h_{s}$, explains the choice of
a conservatively small value of $h =\delta \lambda/2.7$.
%
\subsection{GraF formulae.  The limits and optimum for the FOV width}
\label{graf_formulae}
%
In the absence of the grating, the output image would be a
superposition of all images transmitted in the passbands of the FPI
orders. The grating deviates each order light into a specific angle
according to its wavelength $\lambda_{m}$, and thus spatially sorts the
order images (see Fig.\,\ref{fig:crabe}).

In the output focal plane, the shift between the order images $\Delta w_{f}$ is 
equal to the interorder spectral spacing $\Delta \lambda_{f}$ scaled by the 
reciprocal dispersion $rdisp$ (expressed in $nm/pix$) as follows:
\begin{equation}
\Delta w_{f}~~\textstyle{[pix]} = \frac{\Delta \lambda_{f}}{rdisp}
\label{eq:wf}
\end{equation}
Since the axis of $\lambda$ in the output focal plane is collinear with
the spatial $y$-axis, the value of $\Delta w_{f}$ can be also expressed
in $arcsec$ through the platescale $H$:
\begin{equation}
\Delta w_{f}~~\textstyle{[arcsec]} = P \cdot \frac{\Delta \lambda_{f}}{rdisp}
\label{eq:width}
\end{equation}
Albeit shifted according their wavelength, the order images
superimpose. The confusion is finally avoided by limiting the width of
the entrance FOV to the value $\Delta w_{f}$. The latter, expressed in
$arcsec$, is the confusion free width of the FOV, or simply the free
width (see Fig.\,\ref{fig:crabe}).

The Eq.(\ref{eq:wf}) seems to indicate that the free width $\Delta
w_{f}$ can be increased arbitrarily by increasing the grating spectral
dispersion. However, the extension of the output psf along the
$\lambda$-axis, measured as the FWHM $\delta \phi _{graf, disp}$, also
increases with the dispersion due to collinearity of $\lambda$- and $y$
axes. 

We can write:
\begin{equation}
{\cal{P}}_{graf, disp} = \int_{}{}dw \cdot (M_{FP}(w) \cdot {\cal{P}}_{ao}(w)) 
{\cal{L}}_{gr}(w-\lambda)
\label{eq:psf_graf}
\end{equation}
where ${\cal{P}}_{graf, disp}$ is the psf profile along the
$\lambda$-axis at the GraF output, ${\cal{P}}_{ao}(\lambda)$ is profile
of the monochromatic psf at the output of the adaptive optics,
$M_{FPI}$ is the spectral transmission of the FPI, and
${\cal{L}}_{gr}(w-\lambda)$ is the instrumental spectral profile of the
grating. 

For the corresponding FWHM's, we get :
\begin{equation} 
\delta \phi _{graf,disp} ~~\textstyle{[pix]} \simeq \sqrt{\delta \phi _{ao}^{2} 
+ \left( \frac{ \delta \lambda}{rdisp}\right)^{2} } %
\label{eq:phi_graf}
\end{equation}
which gives the explicit relation of $\delta \phi _{graf, disp}$ and $rdisp$.

\begin{figure} 
	 \centering
	{\includegraphics[height=6cm]{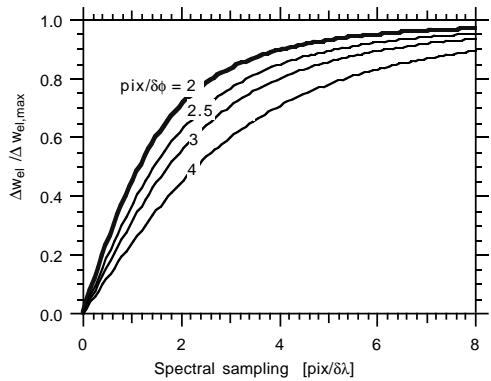}} \caption{The ratio 
	of the free width expressed in number of resolution elements $\Delta 
	w_{el}$ to its asymptotic maximum $\Delta w_{el,max} = {\cal F}$ in 
	function of the spectral (along the grating dispersion) and spatial 
	sampling ratios.}
	\label{fig:free_field}
\end{figure}

To get the insight into the tradeoff guiding the choice of the spectral
dispersion value $rdisp$, let us introduce the dimensionless free width
of FOV, $\Delta w_{el}$, expressed in the number of the spatial
elements of resolution $\delta \phi_{graf, disp}$ as follows:
\begin{equation}
\Delta w_{el} = \frac{\Delta w_{f}}{\delta \phi _{graf,disp}} \simeq 
\frac{\frac{\Delta \lambda_{f}}{rdisp}}{\sqrt{\delta \phi_{ao}^{2} + \left( 
{\frac{\delta \lambda}{rdisp}}\right)^{2}}}
\label{eq:welements}
\end{equation}
From the last equation and Eq.(\ref{eq:finesse}), it follows that when the 
spectral dispersion increases, i.e.\ $rdisp \rightarrow 0$, then:
\begin{equation}
\Delta w_{el} \rightarrow \Delta w_{el, max} = 
\frac {\Delta \lambda_{f} }{\delta \lambda} = {\cal F}
\end{equation}
The finesse ${\cal F}$ sets the asymptotic upper limit to the number of
spatial elements that one can get along the width of the GraF FOV. This
is illustrated in Fig.\,\ref{fig:free_field}, which displays the ratio
of $\Delta w_{el}$ to the possible maximum $\Delta w_{el,max}$ in
function of the spectral sampling ratio along the grating dispersion
$\rho_{\lambda}$. The latter is defined as the number of pixels for one
element of resolution, $\rho_{\lambda} = \delta \lambda /rdsip$. The
curves are computed for several values of the spatial sampling ratio,
$\rho_{s} = \delta \phi _{ao}/H$.

%
One can see that $\Delta w_{el}/\Delta w_{el,max}$ increases rapidly up
to $\rho_{\lambda} \simeq 2.5$ and saturates afterwards. The range of
$rdisp$ and $\Delta w_{f}$ corresponding to $\rho_{\lambda} \simeq 2.5$
can be considered as optimum. Higher dispersions will give only a
slight increase of $\Delta w_{el}/\Delta w_{el,max}$, while the number
of order windows covered by the detector, $M_{wind}$, will decrease
linearly.

The numerical values of the free width $\Delta w_{f}$ chosen for the
instrument are close to the optimum. They and are given in Tab.\
\ref{tab:pars} in $arcsec$ and $pixels$. The corresponding value of
$\Delta w_{el}$, for instance at $\lambda = 2.2~\mu$m, is $\Delta
w_{el} \approx 7$. The other parameters are the platescale $H = 50$
$mas/pix$ (hereafter $mas$ states for $milli-arcseconds$), $rdisp$ of
the 300 mm$^{-1}$ grating, the free width $\Delta w_{f}$ =
1.55{\arcsec} and the measured psf extension $\delta \phi _{graf, disp}
= 0.22{\arcsec}$ (see Sect.\,\ref{data-std}). 

The maximum possible width at this wavelength is $\Delta w_{el,max} =
{\cal F} \approx 16.8$, as calculated from $\Delta \lambda_{f} =
4.75$~nm and the measured spectral resolution $\delta \lambda = 0.3$~nm
of the FPI. The measured ratio $\Delta w_{el}/\Delta w_{el,max} \approx
0.45$ is close to the theoretical value of 0.55 derived from
Fig.\,\ref{fig:free_field} for the used spatial and spectral samplings
ratios respectively $\rho_{s} = 2.6$ and $\rho_{\lambda} = 1.9$.
%
\section{Implementation at the telescope\label{implem}}
%
\subsection{Hardware}
\begin{figure} 
	 \centering
	 {\includegraphics[width=11.5cm]{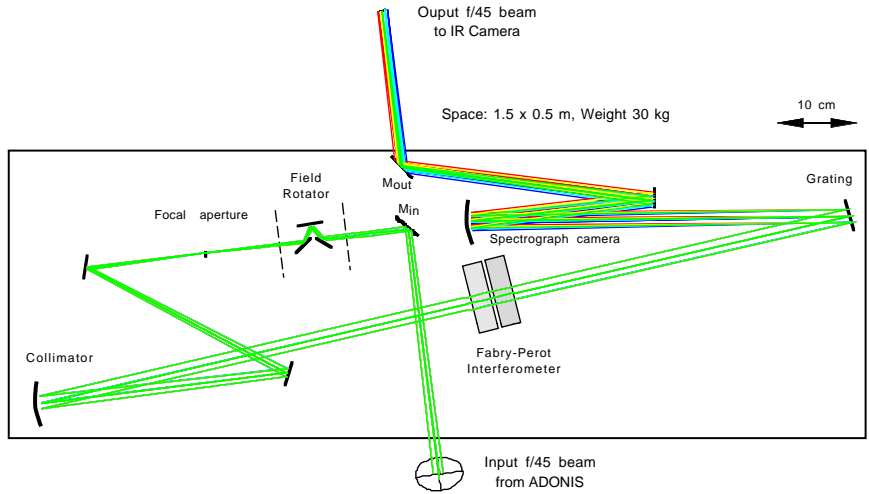}}
	\caption{The GraF optical design.}
	\label{fig:design}
\end{figure}

The implemented spectrograph has been designed to meet the ADONIS imposed 
mechanical constraints, dictating a compact (1.5\,m\,$\times$\,0.5\,m) and light 
(30\,kg) instrument (see Fig.\,\ref{fig:design}).  The operating wavelength 
range from 1.2 $\mu m $ to 2.5 $\mu m $ corresponds to that of the SHARPII+ camera.

The spectrograph was installed at the ADONIS visitor equipment bench 
(\opencite{adonis97}), located between the adaptive optics output and the 
SHARPII+ camera.  The flat mirror M$_{in}$ intercepts the ADONIS f/45 output 
beam and sends it into the spectrograph.  The beam passes firstly through a 
field rotator made of a prism and a flat mirror.  The field rotator of GraF 
allows to vary the slit orientation on the sky, which otherwise would have been 
remained fixed due to ADONIS constraints.
%
%
The FOV is selected by the rectangular aperture located in the ADONIS focal 
plane.  Its width, can be adjusted continuously and with a good precision 
from 0.1{\arcsec} to 15{\arcsec}, and its height is fixed to about 
30{\arcsec}.  After the aperture, the beam is collimated, passes through 
the Fabry-Perot interferometer, is dispersed by the grating, reconfigured 
by the spectrograph camera back to the f/45 beam and sent to the SHARPII+ 
camera.

The instrument can also be used in a direct imaging mode, either by setting the 
grating at the zero order position, or replacing it by a flat mirror mounted at 
the back of the grating (not shown in Fig.\,\ref{fig:design}).  In this 
configuration the FPI is used in the classical scanning mode.

The two flat mirrors M$_{in}$ and M$_{out}$, located on the axis of the 
ADONIS-SHARPII+ camera, are mounted on a dedicated common support, so that they 
can be removed and installed within minutes, allowing a quick change from the 
GraF IFS mode to the regular ADONIS imaging observations.

The motors controlling the focal aperture, the field rotator, the grating 
position, the FPI in- and out- of the beam movements, are operated through the 
GraF dedicated version of the ADOCAM real-time operations software written by 
F.\ Lacombe.  The GraF operation at the telescope thus inherited conveniently 
from the user-friendly interface of the wide-band imaging ADONIS observations, 
and in particular the possibility to launch the command sequences in the batch 
mode.
\subsection{Optical quality}
The necessity to fit the instrument into a reduced room implied adding
5 more flat mirrors in the optical design, further, the necessity of
the field rotator implied adding 3 mirrors more. In total, it makes 11
reflecting surfaces. The loss in transmission is limited by the high
efficiency golden coatings. However, this extra number of optical
surfaces certainly decreased the spectrograph image quality. Estimating
that a mirror is manufactured with $\lambda /10$ precision, and
$\lambda = 0.6$ $\mu m $, the accumulated $rms$ wavefront error should be
about 200\,$nm$, or $\lambda/5$ at the shortest operating wavelength of
$1.2$ $\mu m $, which can be considered as still acceptable.

More importantly, these static aberrations, at least at low and moderate spatial 
frequencies, are corrected by the fine tuning of the AO, so that the 
degradation of the final image quality is negligible as witnessed by the stellar 
images given in Fig.\,\ref {fig:HR8353_d0}.  The AO tuning is done at the 
beginning of each night.
\subsection{Observing modes}
The Table \ref{tab:modes} summarize the observing modes of the GraF 
instrument.  It shows an apparently complex instrument, while in practice 
the change from one mode to another is done in a few dozen of seconds or in 
a few minutes at longest, and can be programmed beforehand using the 
ADONIS/ADOCAM control software scripts.  The availability of the modes of 
direct imaging and of grating spectroscopy (hereafter GS) was very valuable 
during the tests, providing independent and complementary measurements for 
the IFS mode.  Furthermore, the GS mode was extensively used for 
observations of ``linear'' objects like binary stars.

The grating is mounted on a high quality turning support, so that the grating 
angle settings are of a high precision insuring an accurate and stable 
wavelength setting.  This was of special importance for the IFS mode.  Indeed, 
during the FPI scanning the image on the detector undergoes slight shifts along 
the $\lambda$-axis.  We used an automatic procedure compensating these shifts by 
changing the grating angle at each change of channel in order to keep the image 
as precise as possible at the same detector pixels and thus to decrease the 
uncertainties of the flux estimate.
\subsection{Optical parameters}
The spectral resolutions are listed in Tab.\, \ref{tab:resolution}.
%
\begin{table} 
\centering
\begin{tabular}{|c|c|c|c|} \hline
Mode & Focal aperture & FPI & Grating position \\ \hline
Direct Imaging & Open & Out & Zero order, or \\ 
& J \& H: 9{\arcsec} $\times$ 9{\arcsec} & & Flat mirror \\ 
& K: 12.8{\arcsec} $\times$ 12.8{\arcsec} & &  \\ \hline
Grating spectros.  & Open, or narrow slit & Out & chosen $\lambda$ \\
& J \& H: $\simeq$0.2{\arcsec} $\times$ 9{\arcsec} & &  \\
& K: $\simeq$0.2{\arcsec} $\times$ 12.8{\arcsec} & &  \\ \hline
Scanning IFS & Rectangular & In & chosen $\lambda$, compensation \\
& J \& H: $\simeq$0.8{\arcsec} $\times$ 9{\arcsec} & &  of the $\lambda$-shift induced\\
& K: $\simeq$1.5{\arcsec} $\times$ 12.8{\arcsec} & &  by the FPI scan\\ \hline
Scanning FPI & Open & In & Flat mirror \\
imaging spectroscopy & J \& H: 9{\arcsec} $\times$ 9{\arcsec} & & \\
& K: 12.8{\arcsec} $\times$ 12.8{\arcsec} & & \\ \hline
\end{tabular}
\caption{ The summary of the available observing modes.}
\label{tab:modes}
\end{table}
\begin{table} 
\centering
\begin{tabular}{|c|c|c|c|c|} \hline
$\lambda (\mu$m) & {IFS and FPI modes} & \multicolumn{3}{c|} {Grating Spectroscopy} \\
\cline{3-5} %
& & 35 mm$^{-1}$ & 300 $mm^{-1}$ & 600 $mm^{-1}$ \\ \hline
1.25 & 16000 & 200 & 1200 & 2500\\ \hline
1.65 & 10000 & 300 & 1600 & 3500 \\ \hline
2.2 & 7000 & 600 & 4000 & 10000\\ \hline
\end{tabular}
\caption{ Spectral resolving power. For the grating spectroscopic mode, it corresponds to the slit 
width of 0.2{\arcsec}.}
\label{tab:resolution}
\end{table}
The Table\, \ref{tab:pars} gives the main parameters of the GraF IFS mode: the 
reciprocal dispersion $rdisp$, the free FOV width $\Delta w_{pix}$ and $\Delta 
w_{arc}$, and the number of order windows covered by the detector, $M_{ord}$.  
The parameters are listed for the 2 gratings available in the IFS mode and for 
the mostly used platescale $H = 50$ mas/pix.  The parameters for two other 
available platescales, 35 mas/pix and 100 mas/pix, can be obtained by applying 
the scaling factor, except $\Delta w_{arc}$ which does not depend on $H$.

The FPI, fabricated by Queensgate Inc.\ according to our
specifications, has the following parameters: the interorder spectral
spacing 10.1096$\pm$0.0005 $cm^{-1}$, the finesse ${\cal F}$ about 20
depending on the wavelength, the optical transmission in peaks about
85\%. The FPI has two operating spectral ranges, $\lambda \lambda$
$1.2\,-\,2.5$ $\mu m $ and $\lambda \lambda$ $3.8\,-\,4.5$ $\mu m $, due to
the plates coating proposed by E.\ le Coarer. Only the first range was
used with the GraF/ADONIS instrument.
\subsection{Thermal background}
The thermal background of a non-cooled spectrograph is significant at
$\lambda > 2$ $\mu m $. It was the subject of a special analysis during
the design phase. Let us remind that the main source of the thermal
emission is the warm environment of the grating. Indeed, contrary to a
mirror with the bi-univocal correspondence between the incoming and
reflected beams, a grating has more than one order of interference and
consequently more than one input lobes corresponding to the output
beam. It is a well known problem in the visible spectroscopy solved by
limiting the detector passband by an order blocking filter. To reduce
the thermal background, we adopted a similar solution, so that as a
rule the spectroscopy at $\lambda \lambda$ $2 -2.5$ $\mu m $ was done
using the cooled circular variable filter (CVF) (see
\opencite{sharp95}). Its passband $\lambda / \Delta \lambda \simeq 50$
fitted nicely the spectral range covered by the NICMOS detector with
the 300 $mm^{-1}$ grating the most frequently used in the IFS mode.

%
\begin{table} 
\centering
\begin{tabular}{|c|c|c|c|c|c|c|c|c|} \hline
 {\footnotesize $\lambda$} & \multicolumn{4}{|c|}{\footnotesize Grating
300 $mm^{-1}$} &
 \multicolumn{4}{c|}{\footnotesize Grating 600 $mm^{-1}$} \\ \cline{2-9}
& {\footnotesize Disp.} & \multicolumn{2}{c|}{{\footnotesize Free FOV
width}} &
{\footnotesize $M_{ord}$} & {\footnotesize Disp}.  &
\multicolumn{2}{c|}{{\footnotesize
Free FOV width}} & {\footnotesize $M_{ord}$} \\ \cline{3-4} \cline{7-8}
{\footnotesize $\mu m $} & {\footnotesize nm/pix} & %
{\footnotesize arcsec} &{\footnotesize pix} & & {\footnotesize nm/pix} & %
{\footnotesize arcsec} & {\footnotesize pix} & \\ \hline %
{\footnotesize 1.2} & {\footnotesize 0.1655} & %
{\footnotesize 0.43} &{\footnotesize 8.70} & {\footnotesize 29} & {\footnotesize 0.0767} & %
{\footnotesize 0.94} &{\footnotesize 18.8} & {\footnotesize 13} \\ \hline %
{\footnotesize 1.6} & {\footnotesize 0.1621} & %
{\footnotesize 0.79} &{\footnotesize 15.8}& {\footnotesize 16} & {\footnotesize 0.0707} & %
{\footnotesize 1.81} & {\footnotesize 36.2} & {\footnotesize 7} \\ \hline %
{\footnotesize 2.2} & {\footnotesize 0.1558} & %
{\footnotesize 1.55} &{\footnotesize 31.1} & {\footnotesize 8} & {\footnotesize 0.0581} & 
{\footnotesize 4.17} &{\footnotesize 83.3} & {\footnotesize 3} \\ \hline %
{\footnotesize 2.5} & {\footnotesize 0.1520} & 
{\footnotesize 2.06} & {\footnotesize 41.1} & {\footnotesize 6} & {\footnotesize 0.0494} & %
{\footnotesize 6.32} & {\footnotesize 126.5} & {\footnotesize 2} \\ \hline %
\end{tabular}
\caption{ Parameters of the IFS order window for the platescale 50 mas/pix.  The 
FPI interorder spectral spacing $\Delta \sigma_{f} = 10.1096$ $cm^{-1}$.}
\label{tab:pars}
\end{table}
\subsection{Measured sensitivity}
The measured limiting magnitudes $m_{lim}$ in the two main spectroscopic 
modes are listed in Table \ref{tab:mags}.  In the J and H bands, $m_{lim}$ 
is very close to the values derived from the limiting magnitudes of the 
broad-band imaging \cite{adonis_lemignant} and the theoretical GraF 
throughput of 0.64 for the GS mode and of 0.54 for the IFS mode.  The 
throughput was calculated taking the reflexion coefficient of 7 mirrors 
equal to 0.98 each, the transmission of the FPI of 0.85, the efficiency of 
the grating of 0.80, and the transmission of the field rotator of 0.94.  In 
the K band, for the exposures shorter than 1\,min in the GS mode and 5\,min 
in the IFS mode, the noise is dominated by the detector read-out noise 
($\simeq 35$ electrons); above this time, the noise is dominated by that of 
the thermal background.
 
\begin{table} 
\begin{tabular}{|c|c|c|c|c|c|} \hline
{\footnotesize Mode} & {\footnotesize Exp.\ time (min.)} & {\footnotesize J} &
{\footnotesize H} & {\footnotesize K} \\ \hline
{\footnotesize IFS} & {\footnotesize 1} & {\footnotesize 8.5} &
{\footnotesize 9.4} & {\footnotesize 9.3} \\ 
\cline{2-5} & {\footnotesize 5} & {\footnotesize 10.4} &
{\footnotesize 11.1} & {\footnotesize 10.0} \\ \hline
{\footnotesize Grating} & {\footnotesize 5} & {\footnotesize 10.6} &
{\footnotesize 11.4} &
{\footnotesize 10.6} \\ \cline{2-5} {\footnotesize Spec.} & {\footnotesize
30} &
{\footnotesize 12.1} & {\footnotesize 12.9} & {\footnotesize 11.5} \\ \hline
\end{tabular}
\caption{The measured limiting magnitudes for S/N=5 with the grating 300 
$mm^{-1}$.  The integration time for the IFS mode is given per frame.  For 
the grating spectroscopy, the values are given for the slit width of 
0.2{\arcsec}.  The spectral resolutions are those of Table \ref{tab:resolution}.}
\label{tab:mags}
\end{table}
%
\section{Observing procedures and calibrations} \label{observing}
%
We will limit the discussion of the observing procedures to the IFS mode and to 
particularities of the GS mode, other modes being classical.
\subsection{Grating long slit and slitless spectroscopy} \label{gs}
The sharpness of the AO psf ${\cal{P}}$ with $\delta \phi \simeq
0.1{\arcsec}~-~ 0.2{\arcsec}$, much steeper than that of seeing limited
observations (e.g.\ \opencite{racine_psf}), is a new element to take
into account for the GS observations. Setting the slit width close to
the extension of ${\cal{P}}$ as it is usually done for seeing limited
observations, can give rise instrumental effects which could be
neglected so far, like e.g.\ diffraction at the slit border. With the
rapidly variable psf $\cal{P}$, these effects are difficult to
calibrate. Also, the acquisition of the star on the center of the slit
of only $\approx$ 0.1{\arcsec}~-~0.2{\arcsec} wide can take
considerable telescope time.

\begin{figure} 
	 \centering
	 {\includegraphics[width=6cm]{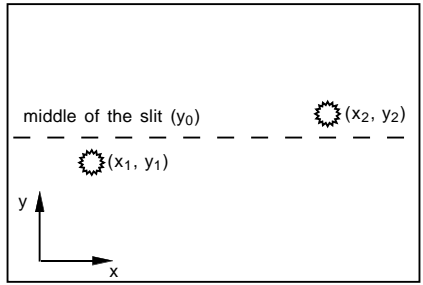}} \\
	\caption{The entrance field of view in the slitless GS-spectroscopic mode 
	operation.  A direct imaging exposure giving the $y$-positions of the stars 
	relative to the middle of the slit allows to adjust the zero's of the 
	wavelength scales of the stellar spectra.}
	\label{fig:open_slit}
\end{figure}

To avoid these losses, we let whenever possible the entrance aperture wide open 
(several arcsec), the sharpness of ${\cal{P}}$ already insuring a well defined 
spectrum.  The important caution in this slitless mode of observations was to 
take a direct image of the field to record the accurate positions of the star(s) 
producing the spectrum (see Fig.\,\ref{fig:open_slit}).  The difference of the 
stellar $y$-positions relative to the middle of the slit provides then the value 
of the translation to be applied to the wavelength calibration obtained as 
usually on a calibration spectral lamp with a narrow slit.
%
%
\subsection{Calibrations} \label{calibrations}
\begin{figure} 
	 \centering
	 {\includegraphics[width=10cm]{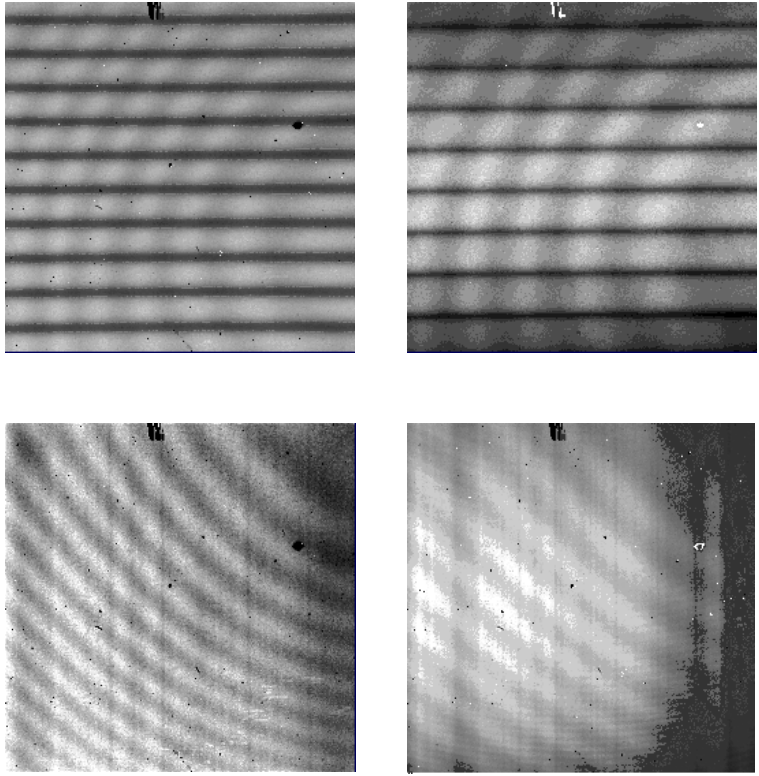}} \\
	\caption{Flat-field frames in the 3D-mode (top) and in the long slit mode (bottom),  
	at $\lambda = 1.6$ $\mu m $ (left), and at $\lambda = 2.2$ $\mu m $ (right). }
	\label{fig:ff}
\end{figure}

%
The main instrumental effects implying the calibration exposures were the usual 
detector pixel-to-pixel response and bias values, and the spectral setting 
parameters.  The complete set of GS data of an object is as follows (with the 
typical global telescope time given in parenthesis):
\begin{itemize}
  \item Scientific long slit or slitless GS frame (10-30 min); %
  \item Image frame of the spectroscopic field of view; Particularly important 
  for slitless GS to define the translation of the wavelength calibration (1
  min);%
  \item Long slit spectrum of the ``flat-field'' produced by a tungsten lamp 
  illuminating the entrance ADONIS pupil; gives the calibration of the detector 
  pixel-to-pixel variations of sensitivity; the slit width 0.2{\arcsec}\,-\,
  0.4{\arcsec} (needs the interruption of the AO servo-loop, 5 min);%
  \item Long slit spectrum of a spectral Ar or Ne lamp, providing the wavelength 
  calibration; the slit width 0.2{\arcsec} (1 min, done during the same 
  interruption of the AO servo-loop as the ``flat-field'' spectrum);%
\end{itemize}
While the wavelength calibration was reliable and better than 0.01\% (see 
Sect.\,\ref{data-std}), the ``flat-field'' calibration of the 
pixel-to-pixel variations of sensitivity turned out to be troublesome.  It 
turned out that the design of the SHARPII+ camera optics and that of the 
NICMOS detector conspired in the way to modulate the overall spectral 
response by the interference fringes formed by the sapphire substrate of 
the detector (see Fig.\,\ref{fig:ff}).  At some wavelengths, the amplitude 
of the fringes was up to 20\% peak-to-peak.  More importantly, the fringe 
pattern could vary, by translation due to probably mechanical flexure, and 
also due to variations of the NICMOS pixels sensitivity.  It was therefore 
mandatory to take the ``flat-field'' spectra very close in telescope 
position and time to the object spectrum.  With all cares taken, the 
residual fringing was still about 1\%, and sometimes worth.

A typical IFS data set is as follows:

\begin{itemize}
  \item The IFS cube of data on the object (from 20 min for exposures of 15\,s
  per frame up to about 1 hour for 60\,s per frame);%
  \item Direct image of the field of view (1 min);%
  \item IFS ``flat-field'' cube (5 - 10 min);%
  \item IFS spectral lamp cube, with the FPI scanning limited to about 10 steps 
  around a prominent spectral line of Ar or Ne, providing the IFS cube wavelength 
  calibration and the measurement of the FPI spectral response (5 min).
\end{itemize}
 
The parallelism of the FPI plates is checked and adjusted in the beginning of 
each night.
%
\section{Data reduction}\label{datared}
%
Although the deconvolution is an integral part of the spectrum extraction 
in the case of scarcely resolved objects (see 
Sect.\,\ref{image_restoration}), a complete discussion of the possible 
strategies in different types of situations is largely out of the scope of 
the present paper.  Here, we will describe only the procedures used to 
extract the spectra of the complex central region of $\eta$ Car object 
chosen to test the instrument performances at the limit of the resolution.
\subsection{Preliminary steps}
They were as follows:
\begin{itemize}
  \item The frame ``cleaning'' for irrelevant pixel values.  It was done 
  using a median filter with the adaptive threshold and the area size, 
  which can run along columns, lines, or square areas according to best 
  final result.  The irrelevant pixel values were replaced by the 
  2-dimensional linear interpolation.  %
 \item Dark frames subtraction; %
 \item Flat-field division.
\end{itemize}
\subsection{Spectrum extraction. GS mode}
Since at the beginning of every run the detector was aligned so that its 
lines were parallel to the the slit.  The grating dispersion is then 
parallel to the detector columns at least in first approximation.  However, 
the parabolic aberration usual to grating spectrographs makes a stellar 
spectrum to be a slightly curved line.  We therefore extracted the spectra 
of stellar sources by fitting the psf ${\cal{P}}$ line by line, with the 
multiplicative factor $A$ (amplitude) and the position across the 
dispersion as the fit parameters.  The amplitude $A$ gives the estimate 
$\hat{S}$ of the flux density.  The psf ${\cal{P}}$ is the average of 
several lines selected in the continuum of an isolated star, if possible 
the studied object, otherwise a reference star off the field. 

If the spectrum was obtained as a slitless (see Sect.\,\ref{gs}), the 
calibration of the line number \textit{vs} the wavelength $\lambda$ is done 
taken into account the position of the star with respect to the slit 
center.  

The spectrum of a standard star in Fig.\,\ref{fig:HR8353_spectrum} (dotted 
line) gives an example of the final result.
\subsection{Spectrum extraction.  IFS mode}
\subsubsection{Isolated stars}
The spectrum extraction for an isolated star is done by fitting a 
two-dimensional normalized psf ${\cal{P}}$ to each order image of a channel 
frame with the amplitude $A$ and the $\{x, y\}$ position as the fit parameters.  
As in the GS case, the amplitude $A$ is used as the estimate $\hat{S}$ of the 
flux density $F$, and after the wavelength calibration the vector $\hat{S}$ 
\textit{vs} $\lambda$ provides then the stellar spectrum 
(Fig.\,\ref{fig:HR8353_spectrum}, solid line).  

The psf ${\cal{P}}$, constructed as the ``shift-and-add'' average of the images 
corresponding to the stellar continuum, is normalized so that the integral of 
the signal over the psf area is equal to one.  This normalization makes the fit 
amplitude insensitive to the variations of the Strehl ratio from one channel 
frame to another.

If each channel frame contains orders corresponding to the stellar 
continuum, then one can monitor the channel-to-channel photometric 
variations.  Obviously, the continuum orders must be selected as free as 
possible from the telluric absorptions.  The channel-to-channel variations 
of the average flux are assumed then to be photometric and give the 
correction factor.  The method was successfully tested on the data of 
HR8353 (see Sect.\,\ref{data-std}).
\subsubsection{Complex objects with the features at the psf scale}
The deconvolution becomes then unavoidable.  Comparison of different strategies 
being out of the scope of the present article, we can give only a short 
description of the recipes used during the data reduction of the test case of 
the $\eta$ Car central region (see Sect.\,\ref{data-etacar}

Firstly, the psf ${\cal{P}}$ is constructed as described for the isolated stars 
cases.  Then the Richardson-Lucy (RL) deconvolution algorithm was run on the 
cube frames.  The RL algorithm provides a sharpened image clarifying the 
structure of the object.  However, the resulting flux distribution is unreliable 
for the quantitative analysis (e.g.\ \opencite{magain98}), and must be found by 
other methods.

So that at the next step we used a simple physical modeling of the studied 
region best fitting the data.  Thus the model for $\eta$ Car consisted of point 
sources at the positions found by RL algorithm. 

The distance between the secondary spots in $eta$ Car, down to 0.11{arcsec}, 
being too close to the resolution limit (0.11{arcsec} at $\lambda$ 1.7$\mu m $), 
the uncertainties of the model fitting turned out to be too important to be used 
for the spectrum extraction.

We used then a specific to $\eta$ Car additional information provided by the 
photocenter spectral dependence in the way known as the differential 
``super-resolution'' (\opencite{beckers82}, \opencite{toko92}).  This allowed 
us to separated the spectrum of the bright star and the remaining secondary 
objects (see for details Sect.\,\ref{data-etacar}).
\subsubsection{Software}
We used the Khoros 2.1 software integration and development environment, which 
provides the visual programming of data flows and an extended library of 
routines suitable for the reduction and analysis of data up to 5 dimensions 
\cite{khoros96}.  When necessary, GraF specific routines were developed and 
added to the Khoros library .
\section{Example and quality of the IFS data}\label{data}
%
We present and discuss below the instrument performances in terms of 
spectroscopic and imaging capabilities measured in two relevant test cases.

The first case is a single standard star.  Its measurement provides a reliable 
estimate of the quality of the spectrum extracted from the IFS data.  We show in 
particular that recording the stellar light simultaneously in several spectral 
``windows'' allows to accurately monitor the channel-to-channel variations of 
flux count estimates.

The second case is the observation of the central 
0.9{\arcsec}$\times$0.9{\arcsec} region of $\eta$ Car, well studied by previous 
research groups.  It provides an excellent test of spectro-imaging ability of 
the instrument in a complex field with the spatial structure at scales down to 
the limit of the angular resolution (about 0.1{\arcsec}) as well as the spectral 
line profile structure at scales close to the limit of the GraF spectral 
resolution (about 10\,000 in this case).
\subsection{Spectrum of a standard star} \label{data-std}
A IFS set of data on a standard star HR8353 B3III, K = 3.31 mag \cite{eso_stds} 
was obtained in November 1997.  The spectral range covering about 40 nm was 
centered on the hydrogen Br-$\gamma$ 2165.55\,nm line.  The FPI interorder 
spectral spacing of about 4.75 nm was covered by 48 FPI channel frames with the 
spectral step between the channels of 9.895$\cdot 10^{-2}$\,nm as measured in 
the FPI order m = 457.  We used the SHARPII+ platescale of 50 mas/pix and the 
grating of 300 $mm^{-1}$ (see Tab.\, \ref{tab:pars} for other parameters).  One 
channel frame records here 8 FPI order images, the total number of spectral 
points is thus 48$\times$8 = 384.  The exposure time per channel was 10\,s.

\begin{figure} [ht]
	 \centering
	{\includegraphics[width=10cm]{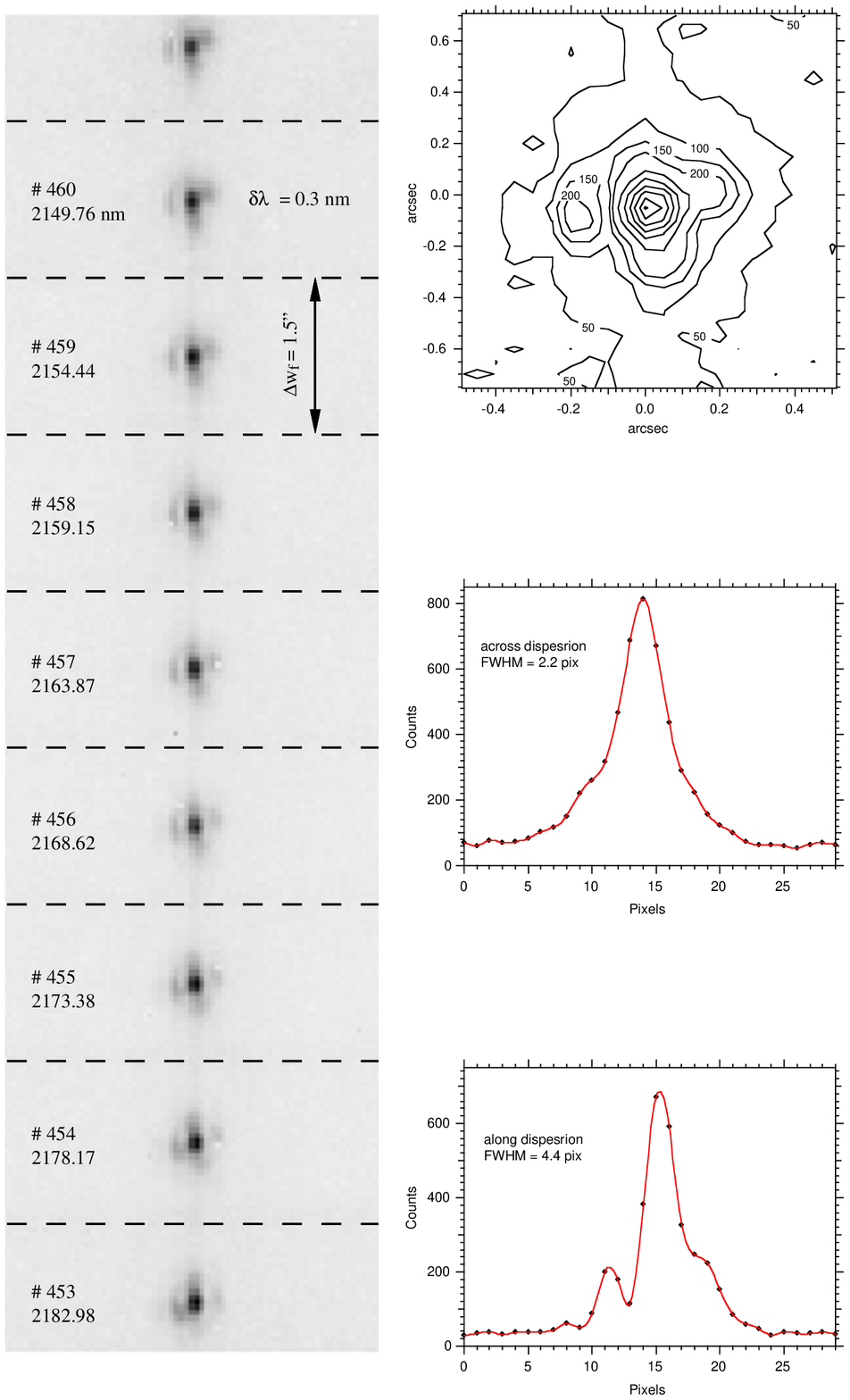}} %
	\caption{A channel frame of the IFS cube of the standard star HR5383.
	\protect\\ %
	\textit{Left panel:} The central strip of the frame corrected for the 
	detector bias and pixel-to-pixel sensitivity variations.  The dashed lines 
	indicate the spatial limits of the FPI order windows.  The spectral passband 
	of window $\delta \lambda$ = 0.3\,nm, or R $\simeq$ 7000.
	\textit{Right panel:} The point-spread function extracted from the order 
	459.  \textit{Top:} The contour map of the stellar image.  The contour 
	decrement is 100 counts above the 200 counts level.  \textit{Middle and 
	bottom}: The counts distribution through the maximum respectively across and 
	along the spectral dispersion; the FWHM $\delta \phi_{x}$ = 2.6 pix 
	(0.13{\arcsec}), and $\delta \phi_{graf,disp}$ = 4.4 pix (0.22{\arcsec}).
	}
	\label{fig:HR8353_d0}
\end{figure}
A strip with the stellar images recorded in the first channel is reproduced in 
Fig.\,\ref{fig:HR8353_d0} (left).  The good image quality of GraF data is 
witnessed by the sharpness of the flux distribution with FWHM $\simeq$ 
0.13{\arcsec} across the grating spectral dispersion, as one would expect for 
the diffraction limited image.  Along the dispersion, FWHM $\simeq$ 
0.22{\arcsec}, as one expects for the convolution product of the diffraction 
limited image with the spectral response of the FPI and the grating (see Sect.\ 
\ref{graf_formulae} and Eq.(\ref{eq:phi_graf})).

\begin{figure} 
	 \centering
	{\includegraphics[width=8cm]{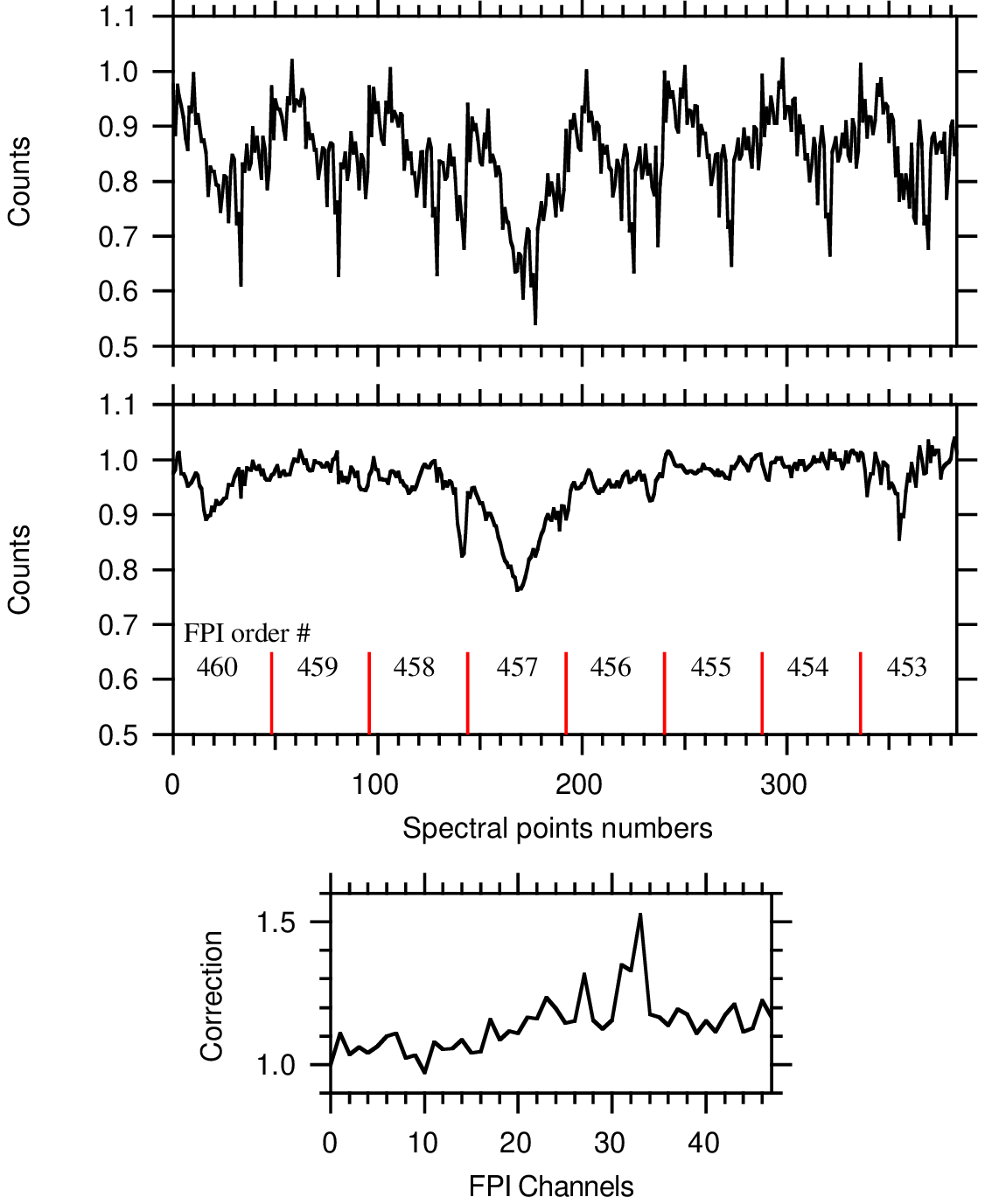}} 
	\caption{Correction for the ``photometric'' variations during the FPI scan.
	\protect\\ %
	\textit{Top and middle:} Flux counts estimate in the channels before (top) 
	and after (middle) the correction for counts variations.  The FPI order 
	numbers and their spectral limits are indicated in the middle plot.  The 
	scan consists of 48 channels.  One channel covers 8 FPI orders, the scan 
	results in 384 spectral points.  The points are numbered in the order of the 
	increasing wavelength.  \textit{Bottom:} The photometric correction vs channel.
	}
	\label{fig:FPI_phot}
\end{figure}
\begin{figure} 
	 \centering
	 {\includegraphics[width=12cm]{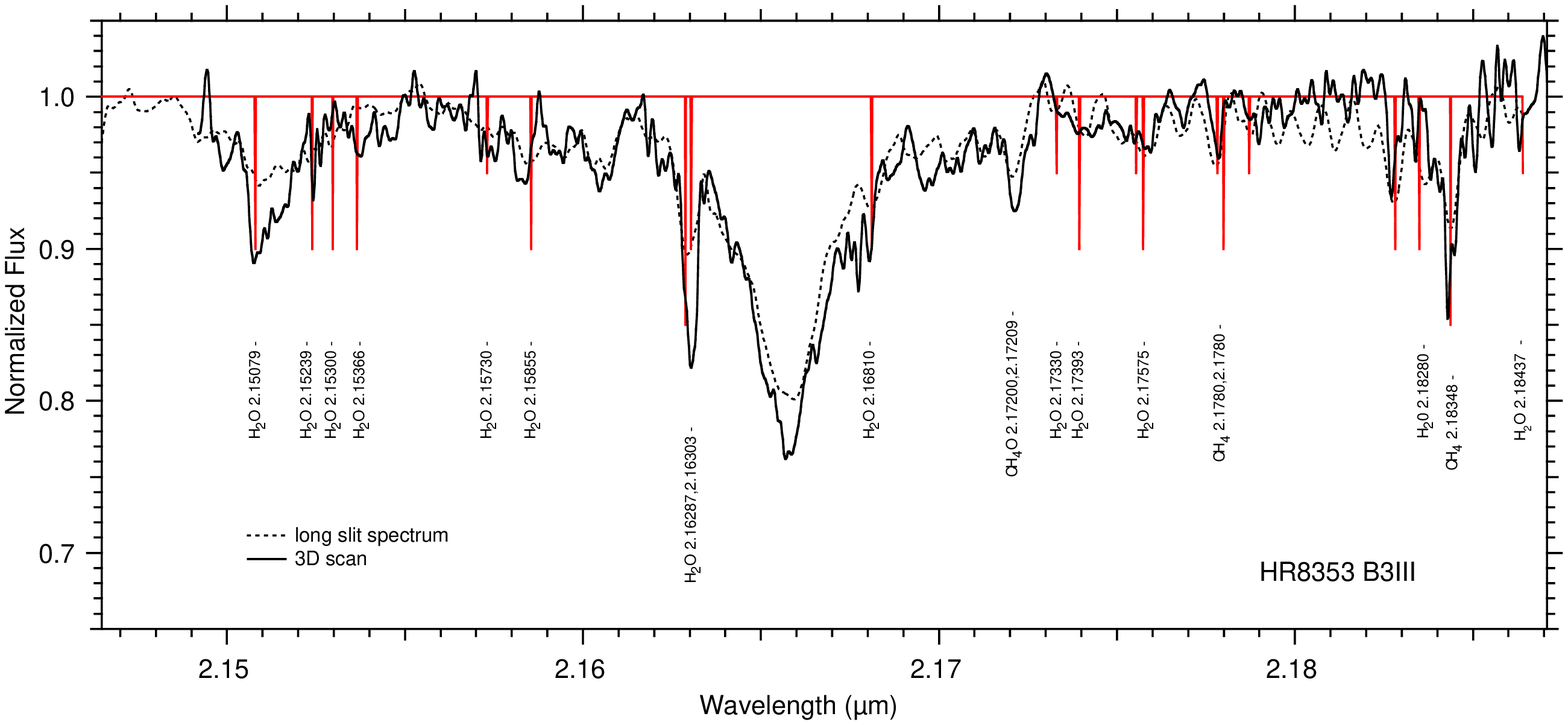}} \\ %
	\caption{Spectrum of a standard star HR8353 extracted from the IFS cube, 
	corrected for the channel-to-channel counts variations and calibrated in 
	wavelength.  The spectrum obtained in the long slit mode is also plotted as 
	the dashed line.  The identified telluric lines are indicated by vertical 
	bars.}
	\label{fig:HR8353_spectrum}
\end{figure}

The data reduction steps after the usual bias subtraction and correction for the 
pixel-to-pixel variations were as follows.  Firstly, the psf ${\cal{P}}$ was 
estimated separately for each channel as the ``shift-and-add'' average of the 
stellar images in the continuum.  Then the flux count in each order window was 
estimated by fitting ${\cal{P}}$ to the stellar images, with the flux count and 
the image photocenter as the free parameters.  The flux estimate in function of 
the spectral point number is plotted in Fig.\,\ref{fig:FPI_phot} (top).  The 
spectral point numbers are in the order of increasing $\lambda$.  One can see 
strong variations of the flux estimate correlated with the channel number.

We selected the order windows free of stellar and telluric spectral lines, and 
imposed the condition such that the total stellar flux in those windows is the 
same for all channels, which is the definition of the stellar continuum.  The 
channel-to-channel variations of the so defined total flux are then ascribed to 
the photometric variations, and used to calculate the correction factor plotted 
in Fig.\,\ref{fig:FPI_phot} (bottom).  Applying the correction factor to the 
extracted spectrum resulted in the rectified spectrum shown in 
Fig.\,\ref{fig:FPI_phot} (middle) and Fig.\,\ref{fig:HR8353_spectrum}. 

In fact, the described channel-to-channel variations were mostly due to the 
method used for the flux estimate, the psf ${\cal{P}}$ fitting, which is 
sensitive to the Strehl ratio, varying with time.  Using other methods, e.g.\ 
integrating on the area of the stellar image gives much better results.  
However, we took profit of this data reduction misadventure to demonstrate the 
possibility of accurate correction of photometric variations.

The wavelength calibration accuracy was estimated by measuring the agreement 
between the expected and the measured positions of the telluric lines.  It was 
excellent with the relative rms error in $\lambda$ being 4$\cdot$10$^{-5}$.

The S/N ratio estimate is biased by the numerous at this sensitivity
faint ($\leq$1\% of the continuum) telluric absorptions due to
H$_{2}$0, CH$_{4}$, etc., mostly identified in
Fig.\,\ref{fig:HR8353_spectrum}, using the tables by
\inlinecite{mohler55}. The lower limit S/N\,$>$\,100 (1 standard
deviation) was measured on 50 points of on the less contaminated
spectral interval starting from $\lambda 21800$\,nm. An additional
estimate was obtained from the GS data recorded as a sequence of 10
identical exposures. The variations of the flux counts from one
exposure to another gave the estimate S/N$\simeq 300$, close to the
expected from the photon statistics for this bright star.
\subsection {Imaging spectroscopy of multiple spots in the $\eta$ Car central region}
\label{data-etacar}
\begin{figure} 
	 \centering
	 {\includegraphics[width=11cm]{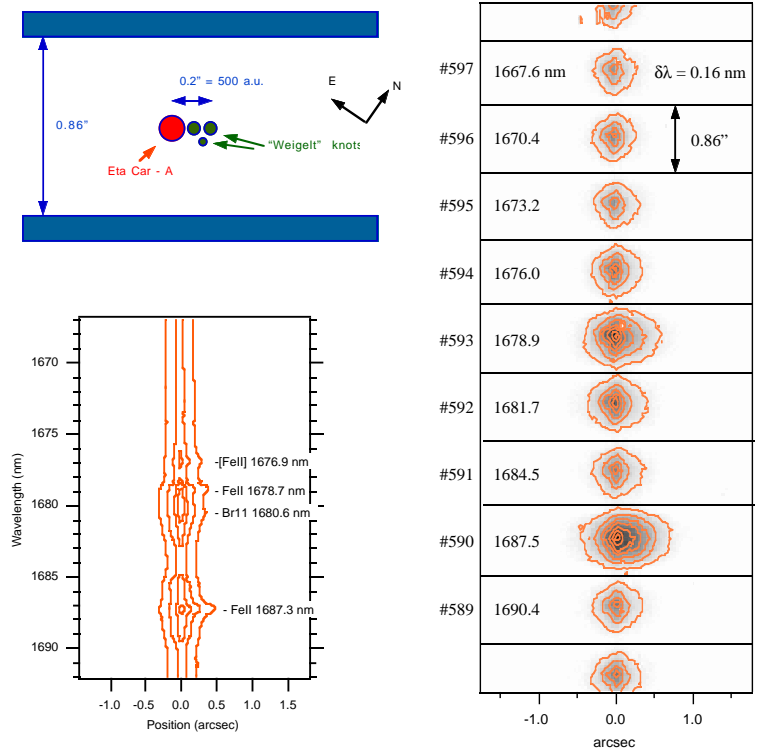}} %
	\caption{Imaging spectroscopy of $\eta$ Car in the 1668-1692\,nm range.  
	\textit{Left top:} The FOV of the IFS cube.  \textit{Left bottom:} The long 
	slit spectrum; the slit width 0.2{\arcsec}, spectral resolution 
	$\simeq$3500.  The contours are at 500, 1500, 2500, 3500 and 4500 counts.  
	\textit{Right} The channel frame 44 of the $\eta$ Car IFS cube.  The FOV and 
	the spectral settings are indicated;
	spectral resolution $\simeq 10000$.}%
%
	\label{fig:eta_3D}
\end{figure}

The $\eta$ Car object is a luminous star loosing mass in giant eruptions (see 
review by \opencite{davidson_araa97}).  The central region contains several 
point-like sources as illustrated in Fig.\,\ref{fig:eta_3D}, the bright spot A 
of $\eta$ Car, and the B, C and D spots at the distance 
0.1{\arcsec}\,-\,0.3{\arcsec} to the NW from the central star 
(\opencite{weigelt86}).  \inlinecite{davidson97}, using HST/GHRS, obtained the 
UV-spectroscopy of the B, C and D spots, showing them to be dense, compact 
slow-moving ejecta emitting in narrow forbidden emission lines of FeII. The 
stellar core emits in broad permitted emission lines formed in the dense fast 
stellar wind (see also \opencite{hamann94} for the ground-based spectroscopy of 
$\eta$ Car with the angular resolution $\delta \phi \approx 5{\arcsec}$ and the 
spectral resolution R $\approx$ 3000 in the near IR, and \opencite{hillier01} for 
the HST/STIS spectrum of $\eta$ Car spot A in the the 164\,-\,1040 nm spectra 
range with $\delta \phi \approx 0.25{\arcsec}$ and R up to 5000).
%
\subsubsection{Observations}
The observations of $\eta$ Car were done on Nov.\,15, 1997 at UT 9:00-9:20.  We 
selected the spectral range 1668-1692\,nm, since it fortunately includes several 
spectral transitions of various excitation, namely the hydrogen Br11 1680.6\,nm, 
[FeII] 1676.9\,nm, FeII 1678.7\,nm, and FeII 1687.3\,nm \cite{hamann94}.  The 
overview of the spectrum is given by the long slit spectroscopy frame 
(Fig.\,\ref{fig:eta_3D}) taken with the slit centered on the spot A and turned 
to the positional angle PA=311{\deg}, between those of the B and C spots at 
PA=340{\deg} and PA=296.5{\deg} respectively \cite{hofmann88}.  The exposure 
time was 15\,s, and the spectral resolution $R\simeq$ 3500.

The IFS data were taken with the platescale 35 mas/pix and the grating 300 
$mm^{-1}$.  The FOV width was 0.86{\arcsec}, and the height 9.0{\arcsec}.  The 
height is fixed by the detector format.  The positional angle of the IFS 
observations was the same as for the GS data, PA=311{\deg}.  The cube consisted 
of 48 channel frames of the FPI. Each channel frame records the FOV in 9 order 
windows with the passband $\delta \lambda = 0.16$\,nm each, or R~=~10\,000.  The 
exposure time per channel was 5\,s.

The instrumental performances in terms of the image quality and deconvolution 
can be illustrated and discussed on the example of one channel frame.  The full 
set of data will be presented and discussed elsewhere together with its 
astrophysical implications.
\subsubsection{Deconvolution}
Already the visual inspection of the cube shows that at the wavelengths of 
[FeII] 1676.9\,nm, FeII 1678.7\,nm and 1687.3\,nm lines the $\eta$ Car image is 
extended.  For the following discussion we selected the channel frame 44, where 
the emission in these lines is at maximum (see Fig.\,\ref{fig:eta_3D}, orders 
593 and 590, corresponding respectively to FeII 1678.7\, nm and 1687.3\,nm).  On 
the other hand, the $\eta$ Car image in the continuum emission, orders 595-597, 
is that of a point-like source, so that we could select one of the orders, 
namely 596, as the psf ${\cal{P}}$ for the deconvolution.

\begin{figure} [ht]
	 \centering
	{\includegraphics[width=11cm]{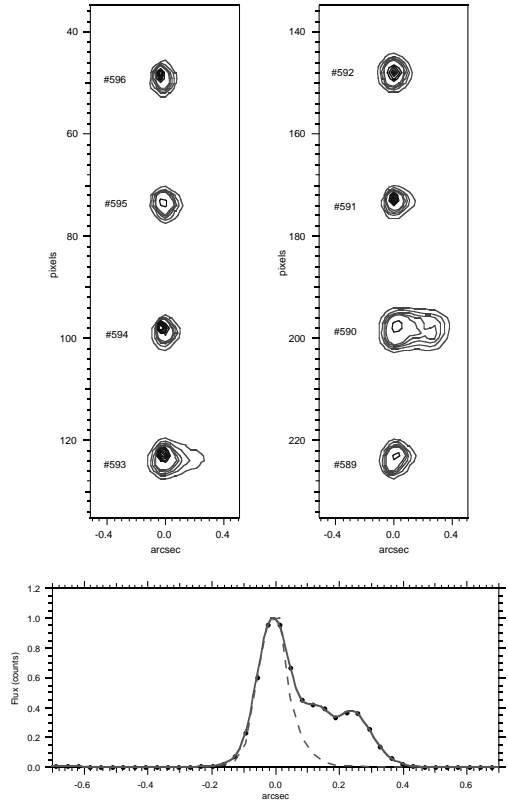}} %
	\caption{\textit{Top:} The deconvolved $\eta$ Car image in the channel frame 
	44.  \textit{Bottom:} The flux profiles across the spectral dispersion in the 
	orders 590 (solid line) and 589 (dashed line).  The profiles are normalized 
	to the maximum value.  The secondary maxima in the order 590 are at 
	0.13{\arcsec} and 0.24{\arcsec} from the maximum flux.  %
	}
	\label{fig:eta_dec}
\end{figure}

The frame was deconvolved using the Richardson-Lucy maximum likelihood algorithm 
(\opencite{lucy74}, \opencite{hook94}), stopped after 50, 200 and 1000 
iterations.  The results obtained with 200 iterations were selected as the best 
compromise between the attained resolution and the level of artifacts.  They are 
displayed in Fig.\,\ref{fig:eta_dec}.
%
%
The regular and sharp appearance of the stellar image in the order 591, produced 
by solely continuum emission and therefore corresponding to the unresolved 
stellar core, witnesses the good quality of the deconvolution and absence of 
artifacts.  The FWHM of this image is 0.1{\arcsec}, very close to the 
diffraction limit of the telescope.

In contrast, the image in the FeII line order 590 shows a complex structure.  
The flux profile across the dispersion (Fig.\,\ref{fig:eta_dec}, bottom) has two 
secondary maxima, at 0.13{\arcsec} and 0.24{\arcsec} from the primary.  This is 
in a good agreement with the values of 0.114{\arcsec} and 0.211{\arcsec} given 
for B and C spots by \inlinecite{hofmann88}, indicating thus the good quality of 
the deconvolved GraF data.  The D spot is not seen probably due to its 
faintness.
\subsubsection{Spatially resolved spectra}
To complete the task of the imaging spectroscopy as it was discussed in
Sect.\,\ref{image_restoration}, we have to solve the image restoration
problem and find the set of estimates ${\hat{S}}(x,y,\lambda)$ of the
flux density distribution. 

Unfortunately, the Richardson-Lucy deconvolution does not restore
properly the flux distribution, enhancing artificially the sharp
sources.

We therefore tried the model fit method. The flux distribution estimate
$\hat{S}$ was modeled as as set of monochromatic functions
${\hat{S}}_{\lambda}(x,y)$ consisting of a primary point-like object
surrounded by 1, 2 or 3 secondary point-like objects. The model
function is then convolved with $\cal{P}$ and we look for $\hat{S}$
giving the best fit to the observed flux distribution
$F_{\lambda}(x,y)$. However, no fit gave satisfaction, either showing
too high residuals (models with 1 and 2 secondaries), or a too shallow
likelihood maximum (models of 3 secondaries) implying highly uncertain
fluxes and positions.

This is not very surprising given that the separation of the A, B and
C spots is very close to the angular resolution limit of the telescope
$\delta \phi \simeq 0.1{\arcsec}$ at this wavelength.

\begin{figure} [ht]
	 \centering
	 {\includegraphics[width=7cm]{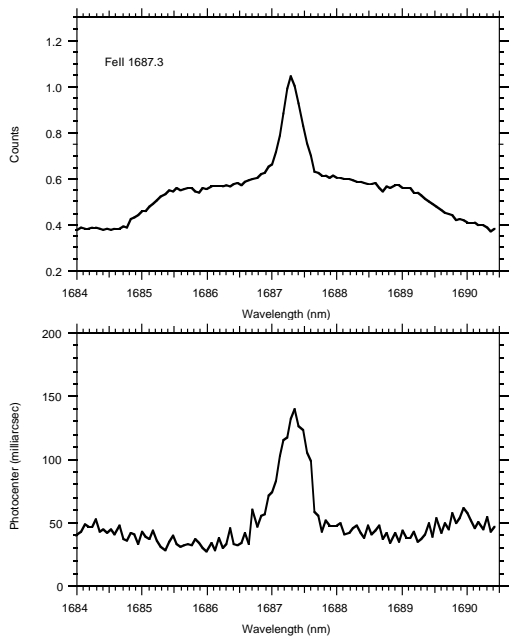}} %
	\caption{\textit{Top: } The profile of FeII 1687.3\,nm in the spectrum of 
	$\eta$ Car.  \textit{Bottom: } The corresponding drift of the photocenter 
	\textit{vs} wavelength.  %
	}
	\label{fig:eta_FeII}
\end{figure}

 The way turned to be fruitful was taking into account the additional
 information provided by the photocenter spectral dependence, following
 the ideas of the differential ``super-resolution'' approach
 (\opencite{beckers82}, \opencite{toko92}). The
 Fig.\,\ref{fig:eta_FeII} displays the integrated spectrum of the whole
 0.9{\arcsec}$\times$0.9{\arcsec} region in the FeII 1687.3\,nm line
 (top), and the photocenter (center of gravity) position across the
 dispersion \textit{vs} wavelength (bottom). The photocenter remains
 constant all along the continuum emission and along the broad wings of
 the FeII line. It deviates up to 150 mas in strong correlation with
 the narrow spectral emission appearing at the top of the broad
 spectral line.

The spectra are separated if we assume a simple model where $\eta$ Car
is composed of 2 objects: the unresolved bright primary (spot A) and an
``ejecta cloud'', which includes both the B and C spots and, possibly,
a halo. Indeed, then the photocenter drift indicates that the the broad
emission comes from the spot A, while the narrow emission well
correlated with the deviation of the photocenter comes from the
``ejecta cloud''.
%

One can derive the parameters of the ``cloud'' emission as follows: 
FWHM\,$\simeq$\,80\,km$\cdot$s$^{-1}$ and 
FWZM\,$\simeq$\,180\,km$\cdot$s$^{-1}$.  The broad emission of the primary spot 
A has FWZM\,$\simeq$\,900\,km$\cdot$s$^{-1}$.  The ratio of the total flux 
emitted in the line by the ``cloud'' to that of the primary is about 0.4.  The 
``cloud'' emission in the FeII 1678.6\,nm line has similar properties.

The presence of emission of two kinds in the near IR lines, one with a large 
velocity dispersion, coming from the stellar wind, and another with a low 
velocity dispersion, has been already reported by \inlinecite{hamann94}, who 
suggested that the first one may arise in a disk surrounding the star.  The GraF 
data obtained at a higher spectral and spatial resolutions showed for the first 
time the velocity resolved profile of the narrow FeII 1676.9\,nm emission and 
clearly indicate that it arises in the ejecta composed of B and C spots to the 
NW from the bright primary A, in a good qualitative agreement with the work of 
\inlinecite{davidson97}.

Clearly, more information can be extracted from the present data, using 
additional spectral lines, etc.  This will be presented in a forthcoming 
article, the discussion of this section being a mere illustration of the 
instrument performances.
\section{Conclusions and prospects}
%
The GraF instrument for imaging spectroscopy based on the Fabry-Perot 
interferometer in the cross-dispersion with a grating was successfully tested 
and operated at the 3.6\,m telescope with the ADONIS adaptive optics, allowing 
to combine the spectral resolution up to 10\,000 with the high angular 
resolution of 0.1{\arcsec}\,-\,0.2{\arcsec} provided by ADONIS. The ability of 
simultaneous imaging in several spectral passbands proved to allow in certain 
cases the simultaneous calibration of the spatial point-spread function 
${\cal{P}}$.  The subsequent deconvolution of data in the crowded field of 
$\eta$ Car is shown to be of a high quality, with the final angular resolution 
close to 0.1{\arcsec} at $\lambda = 1.7~\mu$m, close to the telescope 
diffraction limit, in spite of the fact that at this rather short wavelength the 
adaptive optics correction of the atmosphere induced wavefront perturbations was 
only partial, and the point-spread function was highly variable in time.

The simultaneous mapping of $\eta$ Car with resolutions R up to 10\,000 and 
$\delta \phi \approx$ 0.1{\arcsec} achieved from the ground with 
ADONIS/GraF in the 1.7 $\mu m $ spectral range compares favorably with R up 
to 5\,000 and $\delta \phi \approx$ 0.1{\arcsec} obtained on the same 
object with the space-born instrument HST/STIS in the visible and far-red 
spectral range \cite{hillier01}.

In its practical aspects, the optical concept of the FPI cross-dispersed with a 
grating proved to be compact and versatile, allowing to switch to the long-slit 
spectroscopy, or to direct imaging within a minute.

The potential of the GraF concept has been further proven by the construction of 
new instruments either adopting its principles as the GriF scanning 
integral-field spectrograph for the CFHT (\opencite{grif00}, \opencite{grif02}), 
or developing them further as the Tunable Echelle Imager (\opencite{baldry}).

High quality data were obtained during the period of the scientific use of the 
instrument at the telescope in 1998-2001 on different programmes of stellar 
physics.  They will be a subject of forthcoming papers.
\begin {acknowledgements} %
It is a pleasure to thank people at Laboratoire d'Astrophysique de 
l'Observatoire de Grenoble (LAOG) and elsewhere for discussions and help.  F.\ 
Lacombe (Paris-Meudon Observatory) wrote additional code of the ADOCAM control 
software making possible GraF and SHARPII+ interaction; E.\ Stadler helped with the 
mechanical design; C.\ Tran-Thiet, and G.\ Duvert contributed to the data 
reduction software, R.\ Conan and E.\ Zubia to the numerical simulations, H.\ 
Lanteri and C.\ Aime advised on the data deconvolution, C.\ Perrier, N.\ Hubin, 
J.-L.\ Beuzit, A.\ Tokovinin, R.\ Gredel, S.\ Gilloteau, R.\ Petrov, P.\ 
L\'{e}na, J.\ Melnick, G.\ Monnet have been advising all along the project 
advancement; J.-L.\ Beuzit and A.C.\ Danks helped during the tests at the 
telescope, J.\ Bouvier, T.\ B\"{o}hm, J.\ Eisl\"{o}ffel, C.\ Dougados 
contributed to the scientific case of the GraF proposal document, the ESO La 
Silla staff, in particular J.\ Roucher, E.\ Barrios, J.\ Fluxa, A.\ Gilliotte, 
F.\ Marchis, M.\ Maugis, V.\ Merino, P.\ Prado, A.\ Sanchez, R.\ Tighe, was 
extremely helpful during the GraF tests at the 3.6\,m telescope.  The funding was 
provided by CNRS/INSU/MENSR/Programme National de Haute RŽsolution Angulaire en 
Astronomie, R\'{e}gion Rh\^{o}ne-Alpes, LAOG and Observatoire de Grenoble.  
Scientific collaborations budget was partially supported by the grants from GdR 
``Milieux circumstellaires''.  The telescope technical time was provided by ESO. 
DLM is thankful to ``Soci\'{e}t\'{e} des Amis des Sciences'' for the studentship 
which made possible his stay at Grenoble.
\end {acknowledgements} %
\bibliography{GraF_ExA}
\end{article}
\end{document}